\documentclass{mn}

\usepackage{amsmath}
\usepackage{amssymb}
\usepackage{graphicx,epsfig}

\def\gsim{\;\lower4pt\hbox{${\buildrel\displaystyle >\over\sim}$}\;}
\def\lsim{\;\lower4pt\hbox{${\buildrel\displaystyle <\over\sim}$}\;}
\def\grls{\;\lower4pt\hbox{${\buildrel\displaystyle >\over <}$}\;}

\title[Self-Similar Polytropic Shock Flows]
{Self-Similar Shocks in Polytropic Gas\\
Flows around Star-Forming Regions }

\author[Y.-Q. Lou \& Y. Gao]
{Yu-Qing Lou$^{1,2,3,4}$ and Yang Gao$^{1}$\\
$^1$Physics Department and Tsinghua Centre for
Astrophysics (THCA), Tsinghua University, Beijing 100084, China;\\
\qquad louyq@tsinghua.edu.cn; lou@oddjob.uchicago.edu;
gaoyang-00@mails.tsinghua.edu.cn \\
$^2$Centre de Physique des Particules de Marseille
(CPPM)/Centre National de la Recherche Scientifique \\
\qquad /Institut National de Physique Nucl\'eaire et de Physique
des Particules
et Universit\'e de la M\'editerran\'ee\\
\qquad Aix-Marseille II, 163,
Avenue de Luminy Case 902 F-13288 Marseille, Cedex 09, France;\\
$^3$Department of Astronomy and Astrophysics, The University
of Chicago, 5640 South Ellis Avenue, Chicago, IL 60637 USA;\\
$^4$National Astronomical Observatories of China, Chinese Academy
of Sciences, A20, Datun Road, Beijing 100012, China. }
\date{Accepted 2004... Received 2003...;
in original form 2003}\date{Accepted .
      Received ;
      in original form }

\begin{document}
\maketitle

\begin{abstract}
Self-similar shock solutions in spherically symmetric polytropic
gas flows are constructed and analyzed in contexts of proto-star
formation processes. Among other possible solutions, we model a
similarity shock across the sonic critical curve with an inner
free-fall core collapse and a simultaneous outer expansion of the
extended envelope; the separation or stagnation surface between
these two flow zones travels outwards in a self-similar manner at
a variable speed. One readily obtains accretion shock solutions.
Semi-complete self-similar solutions across the sonic critical
curve either once or twice without shocks can also be constructed.
Features of star formation clouds of our polytropic model include
the mass density scaling in the outer flow zone $\rho\propto
r^{-2/(2-\gamma)}$, the temperature scalings of the inner flow
zone $T\propto r^{-3(\gamma-1)/2}$ and of the outer flow zone
$T\propto r^{-2(\gamma-1)/(2-\gamma)}$, and the variable central
mass accretion rate $\dot{M}=k^{3/2}t^{(3-3\gamma )}m_0/G$ where
$\gamma$ is the polytropic index, $k$ is a constant, $m_0$ is the
core mass, and $G$ is the gravitational constant. Spectral line
profiles characteristic of the `envelope expansion with core
collapse' (EECC) shock solutions are expected. Random magnetic
field permeated in a partially ionized cloud can be incorporated
into this theoretical polytropic model framework. We discuss
briefly our results in context of the oft-observed starless B335
cloud system as an example.
\end{abstract}


\section{INTRODUCTION}

%
In astrophysical and cosmological contexts, the dynamic evolution
of a spherical gas flow has been studied both numerically and
analytically since late 1960s \cite{larson69,penston69}.
Sufficiently far away from initial and boundary conditions, such
flows may gradually evolve into a self-similar phase (Sedov 1959).
Shu (1977) constructed the `expansion wave collapse solution'
(EWCS) in a self-gravitating isothermal gas cloud and advocated
the so-called inside out collapse scenario for forming low-mass
stars \cite{shu77,shu87}. Such a similarity isothermal cloud
collapse process involves a diverging mass density $\rho\propto
r^{-3/2}$ and a diverging radial infall speed $v\propto -r^{-1/2}$
near the collapsing core yet with a $\rho \propto r^{-2}$ far away
in a static envelope characterized by the outer part of a singular
isothermal sphere (SIS). The boundary of the infall region expands
at a constant sound speed. For a more general polytropic gas,
similarity solutions have also been derived and analyzed from
various perspectives
\cite{cheng78,goldreich80,yahil83,suto88,mclaughlin97,fatuzzo04,lou06}.
With effects of unknown energetic processes relegated to the
polytropic index $\gamma$, these polytropic solutions, among
others, become more flexible and versatile in modelling core
collapsing processes in a molecular cloud. Shocks can be
constructed in an isothermal flow
\cite{tsai95,shu02,shenlou04,bianlou05,Yu06}. These similarity
shock solutions contain important realistic features to model
large-scale flow dynamics around star-forming regions. While there
are two- or three-dimensional numerical simulations to model
sub-structures in star-forming regions \cite{truelove98}, we
explore in this paper the full extent of possible one-dimensional
features under spherical symmetry that are extendable to two- or
three-dimensional flow modelling in principle.

Recently, Lou \& Shen (2004) obtained a novel class of
semi-complete isothermal flows, including those referred to as
`envelope expansions with core collapses' (EECC) that may or may
not involve the sonic critical line. Such solutions can model
simultaneous inner collapse and outer flow expansion with or
without shocks and are applied to star-formation regions such as
B335 cloud system \cite{zhou93,shenlou04}. In particular, Shen \&
Lou (2004) emphasized that a cloud envelope, with a central
collapsing core to form a star or a star cluster, may either
expand or contract depending on initial environmental conditions.
Fatuzzo et al. (2004) considered a polytropic gas collapse under
self-gravity in reference to observationally inferred far-away
inflows. They adopted different polytropic indices for equations
of state for an initial static cloud and a cloud in dynamic
evolution in a more general manner. Central mass accretion rates
are also examined as compared with observations (Fatuzzo et al.
2004).
In this paper, we study polytropic gas flow solutions across the
sonic critical curve either with or without shocks, which was not
considered by Fatuzzo et al. (2004). Moreover, semi-complete EECC
solutions crossing the sonic critical curve twice in a polytropic
gas are also constructed as an extension of the isothermal result
of Lou \& Shen (2004). Both shock and EECC features are expected
in clouds with collapsing cores surrounding proto-star forming
regions.
This theoretical development, together with shock modelling (Bian
\& Lou 2005; Yu \& Lou 2005; Yu et al. 2006) and adaptations, may
also bear physical implications for the dynamic evolution phase
with a timescale of $\sim 10^3$yrs linking AGBs and PNe (e.g.,
Balick \& Frank 2002 and extensive references therein)

Observations of star-forming regions have been accumulating over
years. Various optical, infrared, sub-millimeter and more recent
X-ray observations reveal structures on different scales of
hundreds of star-forming regions, a few of which are well observed
\cite{zhou93,shirley02,swift05} and become valuable sources for
checking and testing theoretical model development. By simulating
characteristic spectral line profiles given an underlying model of
dynamic or magnetohydrodynamic (MHD) flows, one can infer
information of a star-forming region by estimating the density and
flow profiles, velocity dispersion, temperature scaling and
central mass accretion rate and so forth.

With these considerations in mind, we present in \S 2 our
polytropic similarity flow model, including the polytropic EECC
and shock solutions. Several aspects of probing star-forming
processes
are discussed in \S 3. Further discussion of our model application
is presented in \S 4. Several details of mathematical derivations
are summarized in Appendices A, B and C for the convenience of
reference.

\section{Self-Similar Polytropic\\
\ \quad Gas Flows }

For the dynamical evolution of spherically symmetric gas flows
under self-gravity and thermal pressure, we use the basic
nonlinear fluid equations in spherical polar coordinates
($r,\theta,\phi$), namely
\begin{equation}
{{\partial\rho}\over{\partial t}}
+{1\over{r^2}}{{\partial}\over{\partial r}}(r^2\rho u)=0\ ,
\label{Equ:mass1}
\end{equation}
\begin{equation}
{{\partial u}\over{\partial t}}+u{{\partial u} \over {\partial
r}}=-{1\over{\rho}}{{\partial p}\over {\partial
r}}-{{GM}\over{r^2}}\ , \label{Equ:force}
\end{equation}
\begin{equation}
{{\partial M}\over{\partial t}} +u{{\partial M} \over{\partial
r}}=0\ , \qquad\qquad {{\partial M}\over{\partial r}} =4\pi
r^{2}\rho\ .\label{Equ:mass2}
\end{equation}
Here mass density $\rho$, thermal gas pressure $p$ and radial flow
speed $u$ are all functions of both radius $r$ and time $t$, and
equations (\ref{Equ:mass1}) and (\ref{Equ:force}) represent mass
and momentum conservations, respectively; $M$ is the total
enclosed mass inside $r$ at a given time $t$ and the first of
equation (\ref{Equ:mass2}) is another form of mass conservation;
the Poisson equation relating gas mass density and the
gravitational potential is automatically satisfied and $G$ is the
universal gravitational constant.

The generalized polytropic relation takes the form of
\begin{equation}
p=K(t)\rho^\gamma\ ,\label{Equ:state}
\end{equation}
where $\gamma$ is the polytropic index and the coefficient $K(t)$
can vary with time $t$. The isothermal case corresponds to
$\gamma=1$ and a constant $K(t)=\kappa$. Equation of state
(\ref{Equ:state}) contains information of some unknown energetic
processes. A time-dependent $K(t)$ leads to different solution
families for $1\lsim\gamma\lsim 2$, while for $p=\kappa
\rho^{\gamma}$, the special cases of $\gamma=4/3$ and $5/3$ should
be handled separately \cite{cheng78,goldreich80,lou06}.

We introduce the self-similarity transformation in the following
form \cite{suto88,lou06}
\begin{eqnarray}
x=r/A(t)\ ,\ M(r,t)=B(t)m(x)\ ,\ u(r,t)=C(t)v(x)\ ,\nonumber\\
\vbox{\vskip 0.5cm} \rho(r,t)=D(t)\alpha(x)\ ,\quad
p(r,t)=E(t)\beta(x)\ ,\quad \label{equ:var}
\end{eqnarray}
where the forms of functions $A(t)$, $B(t)$, $C(t)$, $D(t)$ and
$E(t)$ should be properly chosen. Depending on $x$ only, functions
$\alpha(x)$, $v(x)$, $m(x)$ and $\beta(x)$ are the reduced
density, radial velocity, enclosed mass, and thermal pressure,
respectively. The scale-free independent variable $x$ is defined
following the power-law time dependence assumption $A(t)\propto
t^n$ with $n$ being a constant index; this self-similar
independent variable is thus
\begin{equation}
x=r/(k^{1/2}t^n)\ , \label{equ:varx}
\end{equation}
where $k$ is a constant related to the polytropic sound speed
$c\sim k^{1/2}t^{n-1}$. Here, the scale-free and self-similar
independent variable $x$ is made dimensionless and defined by
$x=r/(ct)$, and the time variation of the polytropic sound speed
$c$
is related to the power-law time dependence $x=r/(k^{1/2}t^n)$.

By taking into account of the following scaling relationships
$u\sim c, \ GM/r\sim u^2,\ M\sim\rho r^3,\ p\sim c^2\rho $ and the
equation of state (\ref{Equ:state}), we can consistently cast the
variables into the following form (see Appendix A for a more
detailed discussion)
\begin{eqnarray}
\rho={\alpha(x)}/{(4\pi Gt^2)}\ ,\
M=k^{3/2}t^{3n-2}m(x)/\big[(3n-2)G\big]\ ,\nonumber\\
\vbox{\vskip 0.5cm} u=k^{1/2}t^{n-1}v(x)\ ,\quad
p=kt^{2n-4}\alpha^{\gamma}(x)/(4\pi G)\ ,\qquad \label{equ:varu}
\end{eqnarray}
where, in reference to equation of state (\ref{Equ:state}),
$K(t)\equiv k(4\pi G)^{\gamma-1}t^{2(\gamma+n-2)}$ is readily
identified,
and the polytropic sound speed defined by
\begin{equation}
c\equiv (\partial p/\partial\rho)^{1/2}
=(k\gamma)^{1/2}t^{n-1}\alpha^{(\gamma-1)/2}\label{equ:soundspeed}
\end{equation}
varies with $t$ and $r$ in general.
By the definition of $K(t)$, $K(t)=\kappa$ becomes a constant
coefficient for $n=2-\gamma$ and this corresponds to a
conventional polytropic gas.

Introducing the self-similar transformation (\ref{equ:varu}) to
nonlinear partial differential equations (\ref{Equ:mass1}),
(\ref{Equ:force}), (\ref{Equ:mass2}), and (\ref{Equ:state}), we
readily obtain two coupled nonlinear self-similar ordinary
differential equations (ODEs) (see Appendix B for a more detailed
derivation), namely
\begin{eqnarray}
{\big[(v-nx)^2-\gamma\alpha^{\gamma-1}\big](dv/dx)}
={(v-nx)^2}\alpha/{(3n-2)}\quad\nonumber\\
+2(v-x)\gamma\alpha^{\gamma-1}/x-(v-nx)(n-1)v\
\label{Equ:function1}
\end{eqnarray}
\noindent and
\begin{eqnarray}
\big[(v-nx)^2-\gamma\alpha^{\gamma-1}\big] \big[d\alpha/(\alpha
dx)\big]=
(n-1)v \qquad\qquad\nonumber\\
-(v-nx)\alpha /(3n-2)-2(v-x)(v-nx)/x\ .\label{Equ:function2}
\end{eqnarray}
In this procedure of derivation, an important result is
$m(x)=\alpha x^2(nx-v)$ (see equation \ref{appen_mres}), giving a
clear physical requirement of $nx-v>0$ for a positive enclosed
mass. All the solutions seriously considered in this paper must
satisfy this necessary requirement. By setting $v=0$ for all
$x>0$, we obtain a static solution from equations
(\ref{Equ:function1}) and (\ref{Equ:function2}), namely
\begin{eqnarray}
v=0\ ,\qquad\qquad
\alpha=\big[{2\gamma(3n-2)}/{n^2}\big]^{{1}/{(2-\gamma)}}
x^{-{2}/{(2-\gamma)}},\nonumber\\
\vbox{\vskip 0.5cm}
m=(2\gamma)^{{1}/{(2-\gamma)}}(3n-2)^{{1}/{(2-\gamma)}}
n^{-{\gamma}/{(2-\gamma)}}x^{{(4-3\gamma)}/{(2-\gamma)}}\
\label{Equ:static}
\end{eqnarray}
(Suto \& Silk 1988;
Fatuzzo et al. 2004), referred to as the singular polytropic
sphere solution (Lou \& Wang 2006).

We consider asymptotic solutions of equations
(\ref{Equ:function1}) and (\ref{Equ:function2}) for $x\rightarrow
0^+$, $x\rightarrow +\infty$ and $x$ near the sonic critical curve
in the next subsection \S 2.1. Semi-complete solutions of the EECC
type and shock solutions are constructed numerically in
subsections \S 2.2 and \S 2.3.




\subsection{Sonic Critical Curves and
Asymptotic Behaviours of Self-Similar Solutions}

Physically, when the travel speed of disturbances relative to the
local flow speed is equal to the local sound speed, a singularity
or a critical point arises. In steady flows of Bondi (1952)
accretion and Parker (1958) solar wind, hydrodynamic solutions are
obtained to go across the sonic critical point smoothly.
Mathematically, we can construct the sonic critical curve by
simultaneously setting the coefficients of $dv/dx$, of
$d\alpha/dx$, and of the right-hand sides of equations
(\ref{Equ:function1}) and (\ref{Equ:function2}) to vanish. The
mathematical form of the sonic critical curve is then specified by
\begin{eqnarray}
nx_0-v_0=\sqrt{\gamma}\alpha_0^{(\gamma-1)/2}\ ,
\quad\nonumber\\
\vbox{\vskip 0.5cm} (n-1)v_0+\frac{nx_0-v_0}{(3n-2)}\alpha_0
-2\frac{(x_0-v_0)(nx_0-v_0)}{x_0}=0\ ,
\quad\label{equ:criticalline}
\end{eqnarray}
where the subscript $0$ in variables $x_0$, $v_0$ and $\alpha_0$
indicates the explicit association with the sonic critical curve.
The first equation in equation (\ref{equ:criticalline}) is
obtained from the coefficients of $dv/dx$ and $d\alpha/dx$. The
second equation in equation (\ref{equ:criticalline}) can be
obtained either from the right-hand side of equation
(\ref{Equ:function2}) or from the right-hand side of equation
(\ref{Equ:function1}), together with the first equation in
equation (\ref{equ:criticalline}). Several sonic critical curves
of different parameters are shown in Figure 1 as examples of
illustration.

To clarify solution behaviours near the sonic critical curve and
to construct solutions across the sonic critical curve, we study
asymptotic solutions (first-order Taylor expansion) of the form
$x=x_0+\delta$, $v=v_0+v_1\delta$ and
$\alpha=\alpha_0+\alpha_1\delta$, where $\delta$ is a small
displacement in $x$. By substituting these expressions into
equations (\ref{Equ:function1}) and (\ref{Equ:function2}), the
leading terms $\alpha_1\equiv d\alpha/dx$ and $v_1\equiv dv/dx$
satisfy the following two equations
\begin{eqnarray}
\sqrt{\gamma}\alpha_0^{(\gamma-3)/2}\alpha_1=v_1-2+2v_0/x_0\ ,
\quad\nonumber\\
\vbox{\vskip 0.5cm}
(\gamma+1)v_1^2+\big[4(\gamma-1)v_0/x_0+n-4\gamma+1\big]v_1
\quad\nonumber\\
+2(2\gamma-1)v_0^2/x_0^2+2\big[\alpha_0/(3n-2)
-2n-4\gamma+4\big]v_0/x_0
\nonumber\\
+(n-2)/(3n-2)\alpha_0+(2n+4\gamma-4)=0\ ,\qquad
\label{equ:criticalasym}
\end{eqnarray}
respectively. Once parameters $n$ and $\gamma$ are specfied and
$x_0$, $v_0$ and $\alpha_0$ are determined from equation
(\ref{equ:criticalline}) for the sonic critical curve, $v_1\equiv
dv/dx$ can be solved from quadratic equation
(\ref{equ:criticalasym}); $\alpha_1\equiv d\alpha/dx$ can then be
obtained from the first equation in equation
(\ref{equ:criticalasym}) accordingly. These eigensolutions in the
vicinity of the sonic critical curve can be utilized to construct
analytically smooth semi-complete solutions across the sonic
critical curve without a shock. As the second equation in equation
(\ref{equ:criticalasym}) is quadratic in $v_1$, along certain
segments of the critical curve, the eigensolutions may not exist.
This situation does arise in constructing solutions shown in
Figure 2.

\begin{figure}
\mbox{\epsfig{figure=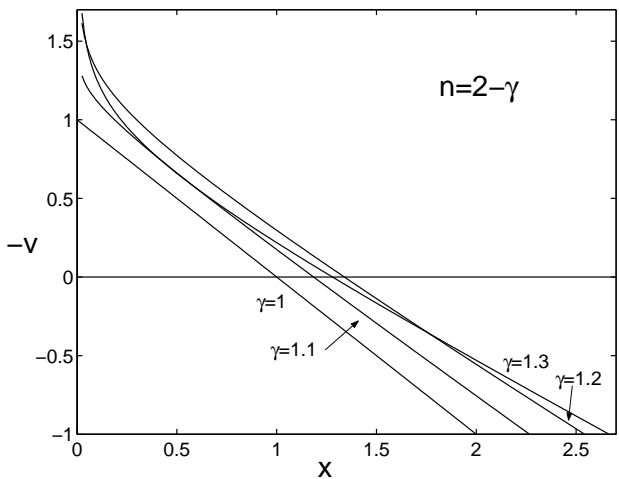,width=\linewidth,clip=}}
\caption{Sonic critical curves with $\gamma=1,\ 1.1,\ 1.2$ and
$1.3$ in the case of $n=2-\gamma$ for a conventional polytropic
gas. The sonic critical curves trace the locus of the sonic
critical points $u=nr/t-c$. The sonic critical curve becomes a
straight line only for $\gamma=1$ corresponding to the isothermal
case. With $\gamma=1.1,\ 1.2$ and $1.3$, the sonic critical curves
diverge for very small $x$. \label{fig0}}
\end{figure}

To leading orders, asymptotic similarity solutions of coupled
nonlinear ODEs (\ref{Equ:function1}) and (\ref{Equ:function2}) for
$x\rightarrow +\infty$ and $x\rightarrow 0^{+}$ are summarized
below. In the limit of $x\rightarrow +\infty$, we have
\begin{eqnarray}
\alpha=Ax^{-{2}/{n}}\ ,\qquad v=-\frac{nA}{(3n-2)}x^{{(n-2)}/{n}}
\qquad\qquad \nonumber\\
\quad\qquad +\frac{2\gamma A^{\gamma-1}}{(2n+2\gamma-3)n}
x^{{(2-2\gamma-n)}/{n}}+Bx^{{(n-1)}/{n}}\ ; \label{Equ:infinity}
\end{eqnarray}
in the limit of $x\rightarrow 0^{+}$, the free-fall solution is
\begin{equation}
v=-\frac{(2m_{0})^{1/2}}{x^{1/2}},\
\alpha=\frac{(3n-2)}{2}\frac{(2m_{0})^{1/2}}{x^{3/2}},\ m=m_{0};
\label{Equ:zero1}
\end{equation}
and in the limit of $x\rightarrow 0^{+}$, the polytropic
counterpart of L-P type solution is
\begin{equation}
x\rightarrow 0^{+}:\quad v={2x}/{3}, \quad \alpha=\alpha^{*},\quad
m={\alpha^{*}x^3}(n-{2}/{3})\ ,\quad\label{Equ:zero2}
\end{equation}
where $A$, $B$, $m_0$ and $\alpha^{*}$ are relevant constants of
integration.
These asymptotic solutions are found by comparing the leading
terms of nonlinear ODEs (\ref{Equ:function1}) and
(\ref{Equ:function2}) for $x\rightarrow +\infty$ and $x\rightarrow
0^{+}$, respectively. For $x\rightarrow 0^{+}$, two different
solutions for the cloud with or without a central core $m_0$ are
considered. If needed, higher-order expansion terms in these
asymptotic solutions can be readily determined.

These leading scalings for $x\rightarrow 0^{+}$ [i.e., inner
solutions for free falls (\ref{Equ:zero1}) and expansions
(\ref{Equ:zero2})] are independent of $\gamma$.
For $x\rightarrow +\infty$, there is a constant speed solution for
$n=1$ corresponding to the isothermal case (Whitworth \& Summers
1985; Lou \& Shen 2004). For $x\rightarrow +\infty$ and
$n=2-\gamma$,
the gas flow Mach number ${\cal M}\equiv {u}/{c}$ remains constant
as shown in equation (\ref{equ:mach}) below.
\begin{eqnarray}
{\cal M}\equiv\frac{u}{c}=\frac{v}{\gamma^{1/2}
\alpha^{(\gamma-1)/2}} =\frac{Bx^{(n-1)/n}+{\cal O}[x^{(n-1)/n}]}
{\gamma^{1/2}A^{(\gamma-1)/2}x^{(1-\gamma)/n}}\nonumber\\
=\frac{B}{\gamma^{1/2}A^{(\gamma-1)/2}}+{\cal O}(x^{-1/n})\ .
\label{equ:mach}
\end{eqnarray}

\subsection{Polytropic EECC Similarity Solutions}\label{eecc}


We numerically solve coupled nonlinear ODEs (\ref{Equ:function1})
and (\ref{Equ:function2}) using available asymptotic solutions as
`boundary conditions'.
Eigensolutions near the sonic critical curve are determined by
equation (\ref{equ:criticalasym}) in order to cross the sonic
critical curve smoothly.
However, for certain segments of the sonic critical curve, such
eigensolutions of $v_1$ and $\alpha_1$ do not exist. It is still
possible to construct shock solutions across these sonic critical
line segments (see \S \ref{shock condition}).

It is simple to construct solutions not crossing the sonic
critical curve by simply integrating from a chosen boundary
condition. For solutions crossing the sonic critical curve
smoothly, matching procedure of Lou \& Shen (2004) is adopted for
constructing semi-complete smooth solutions. That is, for the same
solution, integrated results from the inner and outer boundary
conditions should match with each other at a chosen meeting point.

\begin{figure}
\mbox{\epsfig{figure=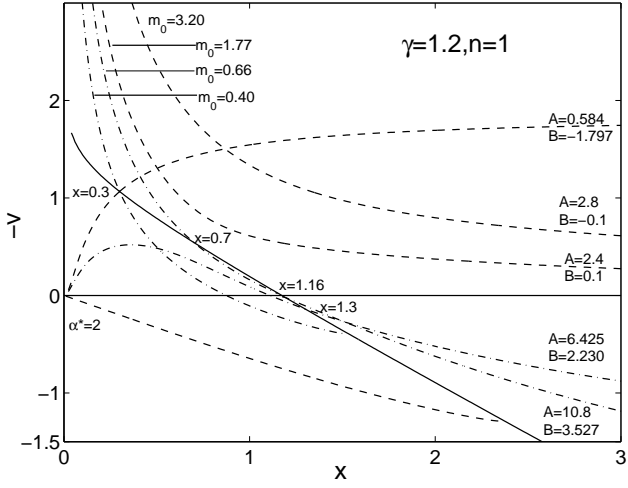,width=\linewidth,clip=}}
\caption{
Three types of polytropic flow solutions for $\gamma=1.2$ and
$n=1$ are constructed by specifying relevant parameters. The solid
line is the sonic critical curve. Dashed curves in the first
quadrant are collapse and inflow solutions.
The incomplete dashed curve in the fourth quadrant
which does not cross the sonic critical line is a counter part of
the L-P solutions.
The dash-dotted curves span two quadrants are EECC solutions.
\label{fig2}}
\end{figure}

Displayed in Figure \ref{fig2} are different types of flow
solutions crossing or not acrossing the sonic critical curve. In
this $-v$ versus $x$ presentation, the first quadrant with $x>0$
and $-v>0$ physically represents collapse or inflow solutions,
while the fourth quadrant with $x>0$ and $-v<0$ physically
represents expansion, wind or outflow solutions. The solid curve
spanning these two quadrants is the sonic critical curve as
determined from equation (\ref{equ:criticalline}). Three types of
flow solutions are collapse/inflow solutions, envelope expansion
with core collapse (EECC) solutions and L-P type solutions are
shown together in Fig. \ref{fig2}.

These different kinds of dynamic solutions originate from
different initial states, carrying sensible physical
interpretations. Those curves diverging at small $x$ have
free-fall features, and the $m_0$ values marked along these curves
represent their core masses. We observe that for a larger core
mass $m_0$, the outer part of the cloud tends to be an inflow. The
solution curve marked with $m_0=3.20$ is connected to an outer
initial state of $B=-0.1$, which represents an inflow (see
equation \ref{Equ:infinity} and note that in this $\gamma=1.2,\
n=1$ case, $B$ is the leading term of $v_{+\infty}$). The flow
curve with $m_0=1.77$ and $B=0.1$ shows a breeze for $x>17.5$,
while the flow curve with $m_0=0.66$ and $B=3.572$ shows an
outflow at large $x$. On the other hand, both the inflow and
outflow initial states at large $x$ may involve an inner state
without a central core (e.g., solution curves with $B=-1.797$ and
$B=2.230$ in Fig. \ref{fig2}). In short, we have the capability of
constructing self-similar flow solutions passing through the sonic
critical curve and making it possible to connect different inner
and outer initial states in all feasible manner. Our polytropic
model framework can accommodate various conditions for the inner
and outer parts of a star-forming cloud.

Solutions spanning both the first and fourth quadrants represent
the EECC feature. We see that both inner core collapse and outflow
features exist in a semi-complete EECC similarity solution. This
may be heuristically viewed as a combination of Bondi (1952)
accretion and Parker (1958) solar wind such that the sonic
critical curve are smoothly crossed twice (Lou \& Shen 2004). We
see also that a L-P type solution and an inner collapse solution
with $m_0=0.40$ do not cross the sonic critical curve because of
the absence of eigensolutions near the sonic critical curve (see
equation \ref{equ:criticalasym} and the relevant discussion).
Nevertheless, this problem can be solved by introducing a shock as
shown in the next subsection. We focus on astrophysical
applications of various EECC solutions in \S 3.

\subsection{Various Similarity Shock Solutions}\label{shock condition}

Shocks are introduced to match solutions from a given outer
boundary condition and a certain solution from inner critical
points or inner boundary conditions. For a polytropic gas, the
shock jump conditions in the shock front reference framework are
\cite{landau59}
\begin{equation}
{\rho_2}/{\rho_1}=\bar{u}_1/\bar{u}_2 =(\gamma+1){\cal
M}_1^2/[(\gamma-1){\cal M}_1^2+2]\ ,\label{shock_v}
\end{equation}
\begin{equation}
{p_2}/{p_1}=2\gamma {\cal M}_1^2/(\gamma+1)
-(\gamma-1)/(\gamma+1)\ ,\label{shock_p}
\end{equation}
\begin{equation}
{T_2}/{T_1}=[2\gamma{\cal M}_1^2-(\gamma-1)][(\gamma-1) {\cal
M}_1^2+2]/[(\gamma+1)^2{\cal M}_1^2]\ ,\label{shock_t}
\end{equation}
where ${\cal M}$ is the flow Mach number in our reference
framework, $\bar{u}$ is the radial gas flow speed
in the shock reference framework and subscripts 1 and 2 denote the
pre-shock (upstream) and post-shock (downstream) sides of the
shock, respectively. The ideal gas law is used here and $T$
represents the gas temperature. These shock conditions are derived
from the mass conservation, the momentum conservation and the
formal entropy conservation along the streamline.

\begin{figure}
\mbox{\epsfig{figure=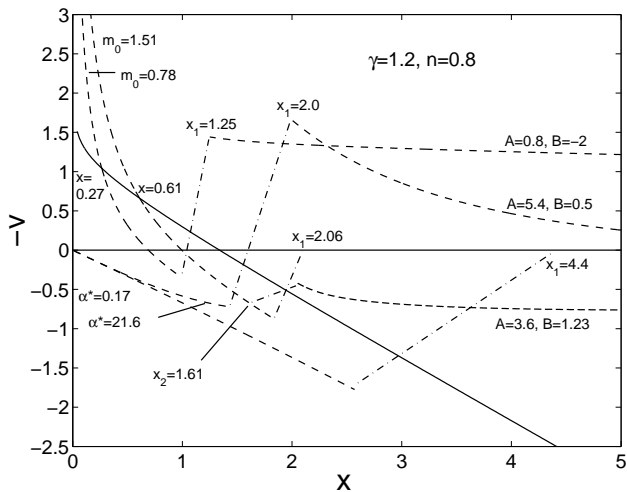,width=\linewidth,clip=}}
\caption{
Three types of shock flows are constructed for $\gamma=1.2$ and
$n=0.8$ with other parameters specified for a conventional
polytropic gas. The solid line spanning the first and fourth
quadrants is the sonic critical curve. Dash-dotted straight lines
connect the pre-shock and post-shock sides (see equation
\ref{equ:shock_discontinuity} and the relevant discussion).
The two dashed curves joining the $-v=0$ axis at $x_1=2.06$ and
$x_1=4.4$ are self-similar shocks running into a static atmosphere
(i.e., the outer part of an singular polytropic sphere). The two
curves with outer initial states characterized by $B=-2$ and
$B=0.5$ show shock discontinuities in the first quadrant; these
are examples of accretion shocks. In contrast, the curve with
$B=1.23$ describes a wind shock.
\label{fig4}}
\end{figure}

For gas flows with a constant polytropic index $\gamma$ for the
equation of state, the sound speed $c$ jumps across the shock
front because of the temperature discontinuity there. That is,
$c_2/c_1=(p_2/\rho_2)^{1/2}/(p_1/\rho_1)^{1/2}=(T_2/T_1)^{1/2}$,
where $c_1$ and $c_2$ are pre-shock (upstream) and post-shock
(downstream) sound speeds, respectively. Our definition of the
self-similar independent variable $x$ is of the scaling $x\sim
r/(ct)$ (see equation \ref{equ:varx} and the relevant discussion).
Due to this jump in the sound speed $c$, independent variable $x$
shifts in the self-similar framework accordingly across the shock
front, namely
\begin{equation}
x_2/x_1=(T_1/T_2)^{1/2}\ .\label{equ:shock_discontinuity}
\end{equation}
Therefore the same physical interface located at $r$ at a given
time $t$ shows a separation across the shock front in our
self-similar presentation because of this shock discontinuity
shown in Figure {\ref{fig4}}.

From equations (\ref{equ:varu}) and (\ref{shock_v}),
we obtain the shock jump conditions in the shock reference
framework in the reduced self-similar form
\begin{equation}
\alpha_2=\frac{\alpha_1(\gamma+1)(v_1-nx_1)^2}
{(\gamma-1)(v_1-nx_1)^2+2\gamma\alpha_1^{\gamma-1}}\ ,
\label{shock1}
\end{equation}
\begin{equation}
v_2=nx_2+\frac{x_2}{x_1(\gamma+1)}\bigg[(\gamma-1)
(v_1-nx_1)+\frac{2\gamma\alpha_1^{\gamma-1}}{v_1-nx_1}\bigg]
\label{shock2}
\end{equation}
(see Appendix C for details of derivation). Because of the entropy
change across the shock front, the constant $k$ is different on
two sides of the shock front and the relation
$(k_2/k_1)^{1/2}=x_1/x_2$ holds.
We determine the post-shock quantities from the pre-shock physical
conditions by applying equations (\ref{shock1}) and
(\ref{shock2}). During this process of shock construction,
equations (\ref{shock_t}) and (\ref{equ:shock_discontinuity}) are
used to determine the post-shock (downstream) independent variable
$x_2$. Thus these shock conditions serve as the condition to cross
the sonic critical curve. Using the matching procedure of Shen \&
Lou (2004), we obtain different classes of shock solutions. That
is, when integrated from an outer boundary condition to a certain
place near the sonic critical curve, the jump conditions
(\ref{shock1}) and (\ref{shock2}) are used to get the post-shock
(downstream) quantities. Simultaneously, another asymptotic
solution from an inner boundary condition is integrated outwards.
When these two solutions match with each other in all quantities
satisfying the shock conditions, a shock solution is then found.

Displayed in Figure \ref{fig4} are different types of shock
solutions with polytropic parameters $\gamma=1.2$ and $n=0.8$.
In this presentation of $-v$ versus $x$, the first quadrant with
$x>0$ and $-v>0$ physically represents collapse or inflow
solutions. While the fourth quadrant with $x>0$ and $-v<0$
physically represents expansion, wind or outflow solutions. The
solid curve spanning these two quadrants is the sonic critical
curve. Dash-dotted straight lines connect the pre-shock and
post-shock sides across the shock. These straight connection lines
are not vertical because the same shock front has different
scalings for the independent variable $x$ on two sides in the
self-similar framework (see equation
\ref{equ:shock_discontinuity}). Three types of shock solutions are
presented together, namely, shocks running into a static
atmosphere, accretion shocks and wind shocks are shown in Figure
\ref{fig4}.

We see that for the curve with the core mass parameter $m_0=1.51$,
different kinds of shocks (shocks running into a static atmosphere
and wind shocks etc.) can be constructed when the shock location
or speed changes. Therefore the shock location is an additional
parameter to join different inner (final) and outer (initial)
conditions. For the same reason, shocks running into a static
atmosphere (e.g., curves with $x_1=2.06$ and $x_1=4.4$) are
constructed in both clouds with and without a central core.

From the two $B=-2$ and $B=0.5$ curves in Figure \ref{fig4}, we
note that an accretion shock can be constructed for a cloud either
with or without a central core mass. A close look at the $B=0.5$
curve shows that this is actually a `breeze' shock solution which
tends to be an outflow at a large $x$ and the flow speed
eventually approaches zero as $x\rightarrow +\infty$ (not shown in
Fig. \ref{fig4} because of scale restrictions). The pure coreless
wind shock with $B=1.23$ is a new addition to the shock solution
family.

These shock solutions show various features for the inner and
outer parts as in Fig. \ref{fig2}. The introduction of shocks
enriches the variety of the semi-complete solution family (because
shock location or speed itself is an additional parameter) and is
qualitatively consistent with observations of star-forming regions
\cite{nisini99}.

\section{Applications to Star-Forming Cloud Cores}

There are several important observational aspects for an isolated
star-forming region: the mass density profile, the temperature
profile, the radial flow speed profile, the core mass accretion
rate, magnetic field, and the angular momentum. In a spherically
symmetric model, we consider and discuss the first five aspects
except the last, and provide estimates of the star-forming cloud
system B335.

Before coming to specific astrophysical applications, we clarify
the physical meaning of similarity solutions passing through the
sonic critical curve. For recent observations of star-forming
regions, the features of inner part and outer edge are well
observed and may serve as initial states for different kinds of
star-formation models. Similarity solutions passing through the
sonic critical curve with or without shocks can join all kinds of
different inner and outer physical states (see Figs. {\ref{fig2}}
and {\ref{fig4}}), while similarity solutions without crossing the
sonic critical curve have much less freedom (Suto \& Silk 1988;
Fatuzzo et al. 2004).
For example, the EECC solutions (Lou \& Shen 2004), most of which
pass through the sonic critical curve, represent a new class of
similarity solutions that are not constructed previously without
encountering the sonic critical curve. Polytropic shock solutions
also form a new family; these shocks connect various inner and
outer physical states and provide theoretical basis for modeling
shock phenomena observed recently \cite{nisini99}.

\subsection{Mass Density and Temperature Profiles
in a Cloud with a Free-Fall Collapsing Core}

By asymptotic solution (\ref{Equ:zero1}) at small $x$, the inner
collapsing core in a free-fall state of a pre-stellar region has a
mass density profile of $\rho\propto r^{-3/2}$. This leading
radial power-law scaling is also predicted by the isothermal model
\cite{shu77,lou04} and appears consistent with observations
\cite{zhou90,andre00}. In other words, this information of inner
mass density scaling alone cannot distinguish between polytropic
and isothermal cloud models.

For the outer part of a pre-stellar region (sufficiently far away
from the collapsing core), the isothermal cloud model gives a mass
density profile $\rho\propto r^{-2}$, while the polytropic cloud
model predicts $\rho\propto r^{-{2}/{n}}$
(see equation \ref{Equ:infinity}). Then for $n=2-\gamma$
corresponding to a constant $K(t)=\kappa$ in a conventional
polytropic gas, the mass density profile for the outer part is
\begin{equation}
\rho\propto r^{-{2}/{(2-\gamma)}}\label{equ:density}
\end{equation}
So $\gamma>1$ means a density exponent smaller than $-2$
corresponding to a steeper density profile, while $\gamma<1$ means
a density exponent between $-1$ and $-2$ corresponding to a less
steep density profile. Earlier observations gave a mass density
profile with an exponent around $-2$ with considerable errors
\cite{zhou90}. Recent observations show that this mass density
exponent for the very outer parts (at radii $r\gsim 15,000$au) of
starless cores is about $-3$ to $-4$ \cite{andre00}, corresponding
to a polytropic index $\gamma\sim 1.3-1.5$. This observational
evidence appears sensible and encouraging for a polytropic model.

Corresponding to the inner mass density profile $\rho\propto
r^{-3/2}$, the radial scaling of thermal temperature $T$ for an
ideal gas in the inner free-fall region would be
\begin{equation}
T\propto p / \rho\propto r^{-3(\gamma-1)/2}\ .\label{equ:temp}
\end{equation}
This inner temperature $T$ would be constant for an isothermal
cloud model with $\gamma=1$. The observed negative index of
temperature scaling \cite{shirley02} calls for $\gamma>1$ in the
inner region, qualitatively similar to the $\gamma$ value inferred
for the outer mass density profile. These radial scalings of cloud
temperature are important features to constrain a polytropic cloud
model by equations (\ref{equ:density}) and (\ref{equ:temp}).

For the reference of observational diagnostics, the gas
temperature sufficiently far away (i.e., large $r$) from the
collapsing core is given either by a constant $T$
for an isothermal cloud model or by
\begin{equation}
T\propto p/\rho\propto r^{-2(\gamma-1)/(2-\gamma)}\
\label{equ:tempfarpol}
\end{equation}
for a conventional polytropic cloud model. Of course, the complete
temperature profile in a conventional polytropic gas can be
readily computed for observational comparison. The key message
here is that independent determination of cloud temperature is
extremely valuable for distinguishing models for cloud dynamics.

\subsection{Radial Gas Flow Speed Profiles}

Outflow features in star-forming regions are well documented in
recent years. In spite of the large-scale bipolar morphology of
these outflows, we are mainly interested in core regions of these
outflow sources with free-fall collapse or infall characteristics
\cite{andre00,wu04,swift05} that are grossly in accord with our
shock scenario of EECC.

Collapses and outflows were thought to happen in different stages
during a star formation process, and they were considered
separately \cite{shu77,yahil83}. However, at least $\sim 12\%$ of
identified outflows show characteristics of collapses or infalls
\cite{wu04}. Note that infall characteristics are more difficult
to identify than those of an outflow due to the complexity of
observed spectral lines and in the analysis of line profiles. One
would expect that more infall motions can be identified in outflow
sources in the future.

Our polytropic EECC shock solutions provide underlying dynamic
models for simulating spectral line profiles, during a certain
epoch of a star formation cloud.
The most expected line profile features are those from EECC
solutions, that is, inflow and outflow within and outside a
certain radius $r$. Double peak spectral line features
representing outflow will be observed away from the core, and
combined double peak line features from both outflow and inflow
will be seen around collapsing cores.

Using an isothermal model for self-similar cloud dynamics (Lou \&
Shen 2004), Shen \& Lou (2004) emphasized the notion that a
molecular cloud with a free-fall collapsing core at the centre can
have various possible forms of motion sufficiently far away from
the collapsing core, depending on the evolutionary history of the
cloud. Such far away radial flows of a cloud may be characterized
by a wind, a breeze, an outer part of a static singular isothermal
sphere (SIS), a contraction, or a inflow. Furthermore, all these
possible radial flow profiles can involve shocks across the sonic
critical curve (Tsai \& Hsu 1995; Shu et al. 2002; Shen \& Lou
2004; Bian \& Lou 2005). And this can give rise to a wide range
for the central mass accretion rate to be discussed presently in
the next subsection. All these important features in an isothermal
cloud model can be further incorporated into a polytropic cloud
model.

In reference to extensive cloud observations, Fatuzzo et al.
(2004) noted a ubiquitous feature of far away inflows and/or
contractions in clouds with central free-fall collapsing cores.
This differs from the static SIS envelope as initially introduced
by Shu (1977) for the EWCS for star formation. The polytropic
cloud model of Fatuzzo et al. (2004) also allows the polytropic
indices to be different for an initially static cloud and for the
subsequent evolution of a dynamic cloud. Nevertheless, Fatuzzo et
al. (2004) did not consider self-similar polytropic flow solutions
that cross the sonic critical curve smoothly and that involve
shocks. Our model analysis here opens up these two important
possibilities in a unified theoretical framework. Moreover, we
emphasize again that in a polytropic cloud model, a cloud with a
central free-fall core collapse can be simultaneously
characterized by a far-away polytropic wind ($B\neq 0$ in equation
\ref{Equ:infinity}) or a far-away polytropic breeze ($B=0$ and
$A\neq 0$ in equation \ref{Equ:infinity}) with or without shocks.
Gas flows with central core collapse and far-away outflow or
breeze are common in star-forming regions \cite{wu04}, indicating
the capacity of our solutions passing through the sonic critical
curve.

It is highly desirable to synthesize spectral line profiles using
radial flow speed and mass density profiles from these underlying
self-similar dynamic models for developing radiative diagnostics
and for extensive observational comparisons.

\subsection{Variable Central Mass Accretion Rate}

The core mass accretion rate from equations (\ref{equ:varx}),
(\ref{equ:varu}) and (\ref{Equ:zero1}) is given by
$\dot{M}=k^{3/2}t^{(3n-3)}m_0/G$. For $n=2-\gamma$ corresponding
to a constant $K(t)=\kappa$ of a conventional polytropic gas, the
core mass accretion rate appears as
\begin{equation}
\dot{M}=k^{3/2}t^{(3-3\gamma )}m_0/G\ \label{accretion rate1}
\end{equation}
(see also equation 31 of Fatuzzo et al. 2004; the difference in a
numerical factor is related to a different definition of
notations), which is constant in the isothermal case of $\gamma=1$
\cite{shu77}.
A constant $\dot{M}$ is inconsistent with observations within a
formation timescale of $\sim 10^5-10^6$yr \cite{andre00}. The
polytropic analysis does offer a possible leeway to model a
variable $\dot{M}$ in time $t$. For a fixed $m_0$, there are two
parameters $k$ and $\gamma$ to adjust. Here $k$ is related to the
sound speed $c$ (see equation \ref{equ:varx}) and $\gamma$ can be
inferred observationally from the radial mass density profile as
discussed in \S 3.1. For various shock solutions, the $m_0$
parameter can vary in a wide range. Although not easily done, it
is crucial to identify the evolution phase of a dynamical cloud
system and estimate or guess $t$ accordingly as discussed
presently.
%

For $\gamma<1$ and $\gamma>1$ in equation (\ref{accretion rate1}),
the core mass accretion rate $\dot{M}$ increases and decreases
with time $t$, respectively. Observations of star-forming systems
\cite{andre00} indicate that mass accretion rates generally
decrease in time $t$. This appears consistent qualitatively with
$\gamma>1$ for most cloud systems with free-fall collapsing cores.
This inference of $\gamma>1$ also agrees with the interpretation
of steeper outer density profiles and the inner temperature
power-law scaling with a negative index (see \S 3.1). For shock
flows, the core mass accretion rate varies with shock locations
and thus shock speeds \cite{shenlou04,bianlou05}. This is because
the value of $m_0$ is different for different shock locations.

The reason why $\gamma>1$ or $\gamma<1$ induce different $\dot{M}$
is as follows. For $n=2-\gamma$ in equation (\ref{equ:varx}), a
physical interface for a fixed $x$ travels with increasing speed
for $\gamma<1$. When this interface is set between the inflow and
outflow regions at $v=0$, we have an increase of infall mass
$\dot{M}\propto\rho r^2dr/dt$. For a fixed $x$, the product $\rho
r^2$ also increases with time $t$ for $\gamma<1$, then the
increasing interface speed $dr/dt$ leads to increasing mass infall
rate. For $\gamma>1$, the interface travels at a decreasing speed
and the product $\rho r^2$ also decreases, thus the mass infall
rate decreases. When we set the interface at the shock location,
the conclusion is that $\gamma<1$ and $\gamma>1$ lead to
acceleration and deceleration of shocks, respectively.

\subsection{Star-Forming Cloud System B335}

The cloud system B335 is a well-studied protostellar collapse
candidate that has been observed through molecular lines
\cite{zhou90,harvey03,evans05}. Data fitting and analysis of
$H_2CO$ and $CS$ molecular transition lines agrees well with the
inside-out collapse model (Shu 1977; Shu et al. 1987), yet the
angular resolution of observations is still not good enough
\cite{zhou93}. Recent observations of Institut de Radio Astronomie
Millim\'etrique, Plateau de Bure Interferometer (IRAM PdBI) give
several new results on this protostellar core \cite{harvey03}.
First, a compact component with a circumstellar disc was
identified. Secondly, the inner mass density profile is
$\rho\propto r^{-p}$ with $p=1.55\pm 0.04$ in a certain region
$r<R_0$. Thirdly, another valuable information is the gas
temperature profile of $T\propto r^{-0.4}$ in a certain region
$r<R_T$. Here, we tentatively take $R_0\sim R_T\sim 5000$au.

The inner mass density profile $\rho\propto r^{-p}$ with $p=1.55
\pm 0.04$ agrees with a free-fall core collapse ($p=1.5$).
Both polytropic and isothermal models show this scaling result.
The outer density profile ($3,500-25,000$au) also shows a power
law with an index $p=1.91\pm0.07$. This result may be regarded as
consistent with the isothermal model with $p=2$, while the
polytropic model can fit this result more accurately if we set
$\gamma=0.95$ in the outer part. One difficulty for the isothermal
model is that the shallow slope ($p\sim1$) of density profile at
the edge of collapse region is still not observed \cite{harvey03}.
But no shallow slope is necessary in the polytropic cloud model.

The temperature profile is $T\propto r^{-0.4}$ for $r<R_T=5000$au,
that is $T=10K$ at $r>R_T$ and increase to $T\simeq50K$ at
$r\simeq100$au \cite{harvey03}.
As observations show that $\rho\propto r^{-1.55}$ near the core,
we need to set $\gamma=1.26$ in the inner region to fit $T\propto
r^{-0.4}$ (see equation \ref{equ:temp}). While one might use the
isothermal model and assume different temperatures in different
cloud regions, the polytropic cloud  model with a temperature
variation appears more sensible.


From the outer mass density profile and the inner temperature
scaling of B335 system, we infer $\gamma=0.95$ for $3500$au $\lsim
r\lsim 25,000$au and $\gamma=1.26$ for $r\lsim 5000$au. This
difference of $\gamma$ shows that $\gamma$ is not a constant at
least for certain cloud systems with core collapses. Only when the
outer mass density profile is steeper than $\rho\propto r^{-2}$,
can $\gamma$ become a constant.

Outflow and shock phenomena in B335 are also reported in recent
years \cite{nisini99}. While previous isothermal solutions
\cite{shu77} or polytropic solutions without crossing the sonic
critical curve (Fatuzzo et al. 2004)
cannot fit these observed conditions, these physical conditions
can be related and constructed using an EECC shock curve passing
through the sonic critical curve. Thus, this cloud system may be
modelled as EECC self-similar shock solutions in a polytropic gas.

\subsection{Magnetic Fields in Star-Forming Clouds}

When a cloud is partially ionized, the magnetic field permeated in
the cloud can exert an influence on the MHD evolution of the cloud
system. Physically, magnetic field directly interacts with the
charged particles which in turn interact with neutral particles
through sufficiently frequent collisions. If the MHD timescale of
interest is much longer than collision timescales between
electrically charged and neutral particles, we may ignore the
effect of ambipolar diffusion and adopt a MHD approximation to
describe large-scale flows of a magnetized cloud. Moreover, the
presence of magnetic field also leads to useful radiative
diagnostics. For example, the presence of relativistic electrons
produced by shocks or proto-stellar activities and of magnetic
field together can give rise to cyclotron and synchrotron
emissions.

Depending on the evolutionary history, magnetic field in a
collapsing cloud under self-gravity may possess a variety of gross
geometries (e.g., Chandrasekhar \& Fermi 1953; Chandrasekhar 1954;
Strittmatter 1966; Bisnovatyi-Kogan, Ruzmaikin, \& Sunyaev 1973;
Mestel 1986; Rees 1987; Lou 1996). We outline below a possible
scenario involving magnetic field to pave the way for further
investigations. It is perceived that a partially ionized gas cloud
under self-gravity is permeated with a completely random magnetic
field (e.g., Zel'dovich \& Novikov 1971). For such a random
magnetic field, we envision a random `thread ball' scenario in a
vast spatial volume of gas medium. A magnetic field line follows
the `thread' meandering within a thin spherical `layer' in space
in a random manner. Strictly speaking, there is always a random
weak radial magnetic field component such that random magnetic
field lines in adjacent `layers' are actually connected throughout
in space. By taking a large-scale ensemble average of such a
magnetized gas cloud, we are then left with packed `layers' of
random magnetic field components dominantly transverse to the
radial direction. Over large scales, the MHD collapse and flow in
such a magnetized cloud are approximated as quasi-spherically
symmetric (Yu \& Lou 2006; Yu, Lou, Bian \& Wu 2006; Wang \& Lou
2006).

In addition to the thermal gas pressure force and self-gravity in
the cloud, the magnetic Lorentz force consisting of magnetic
pressure and tension forces will come into play for a sufficiently
strong magnetic field. MHD generalizations of isothermal
hydrodynamic self-similar solutions are readily available (Yu et
al. 2006). For example, for the MHD free-fall asymptotic solution
towards the core, the magnetic field strength diverges yet the
self-gravity remains overwhelming over the thermal pressure force
and the Lorentz force combined. In such an MHD core collapse
process, a strongly magnetized environment can be created and
sustained surrounding a proto-stellar core. Accretion and rebound
MHD shocks can emerge (Bian \& Lou 2005; Yu et al. 2006; Wang \&
Lou 2006). Near the very centre, various magnetic activities,
persistent or sporadic, are naturally expected to break the
quasi-spherical symmetry as well as the self-similarity. If a
cloud possesses an initial angular momentum, then such a core
collapse under self-gravity will lead to a rapid gas rotation and
a disc formation surrounding the proto-stellar core; in such a
strongly magnetized environment, the coupling between the magnetic
field and disc rotation can give rise to collimated outflows such
as jets or winds (e.g., Lovelace 1976; Blandford \& Payne 1982;
Shu et al. 1997; Ferrari 1998). This type of MHD collapse
processes may lead to intensely magnetized proto-stellar objects.

\section{Discussion}

For star-forming cloud systems, the outer mass density profile,
the inner and outer temperature profiles, and the core mass
accretion rate are crucial to test or constrain a polytropic cloud
model. Given a self-similar dynamic profile in a polytropic gas,
spectral line profiles can be simulated. These spectral line
profiles are very important to show different evolution epochs in
forming proto-stars. In addition to forming proto-stars, the
variable core mass accretion rate $\dot M$ by equation
(\ref{accretion rate1}) also bears physical implications for the
dynamic evolution phase with a timescale of $\sim 10^3$yrs linking
AGBs and PNe \cite{balickfrank02}. Further observations and
analysis are needed especially for post-PNe of different ages. We
also expect adaptations of our polytropic shock model to a certain
epoch of supernova explosions (Lou \& Wang 2006), with radiation
and neutrino pressures also taken into account.

Shock phenomena are common in forming proto-stars, formation and
evolution of PNe, and supernova explosions; they are important for
the gas heating, the production of high-energy photons, the
synthesis of heavy elements, and the acceleration of cosmic ray
particles. In these contexts as well as others, our polytropic
shock model has wide applicability. Properties of polytropic
shocks should be explored further to include radiation processes,
magnetic field effects (Yu et al. 2006; Wang \& Lou 2006) and
atomic/nuclear processes associated with shocks.

Radiative transfer processes will be incorporated into our model
to make the underlying dynamic processes `visible'. Spectral line
profiles of certain atoms and molecules can reveal profiles of
mass density, flow velocity and gas temperature. The EECC profiles
and shocks are the most important features. Certain spectral line
features derived from various underlying dynamic solutions should
be compared with observations. Recent and further high-resolution
spectral IR and microwave telescopes will surely stimulate the
development of dynamic study of star-formation,
formation/evolution of PNe, SNe and other relevant astrophysical
processes.




\section*{Acknowledgments}
This work was supported in part by the ASCI Center for
Astrophysical Thermonuclear Flashes at the University of Chicago,
by the Special Funds for Major State Basic Science Research
Projects of China, by the Tsinghua Center for Astrophysics (THCA),
by the Collaborative Research Fund from the National Science
Foundation of China (NSFC)
for Young Outstanding Overseas Chinese Scholars (NSFC 10028306) at
the NAOC, CAS
by the NSFC grants 10373009 and 10533020 at Tsinghua University,
and by the Specialized Research Fund for the Doctoral Program of
Higher Education
20050003088 and the Yangtze Endowment from the Ministry of
Education at Tsinghua University. The hospitalities of the
Mullard Space Science Laboratory of University College London, of
Astronomy and Physics Department at University of St. Andrews,
Scotland, United Kingdom, and of Centre de Physique des Particules
de Marseille (CPPM/IN2P3/CNRS) et Universit\'e de la
M\'editerran\'ee Aix-Marseille II, France are also gratefully
acknowledged. Affiliated institutions of Y-QL share this
contribution.

\appendix

\section[]{Derivation of Self-Similar Transformation}

We here provide a heuristic derivation of self-similar
transformation (\ref{equ:varu}). In the end, it is the self
consistency counts.

The independent self-similar variable is [see eq.
(\ref{equ:varx})]
\begin{equation}
x=r/(k^{1/2}t^n \label{appen_x})
\end{equation}
and thus the polytropic sound speed $c$ scales as $\sim k^{1/2}
t^{n-1}$. As the radial flow speed $u$ and the polytropic sound
speed $c$ are expected to scale similarly, i.e., $u\sim c$, the
radial flow speed takes the following form
\begin{equation}
u=C(t)v(x)=k^{1/2}t^{n-1}v(x)\ .\label{appen_u}
\end{equation}

From the scaling relationships $M\propto\rho r^3$ and $GM/r^2 \sim
u^2/r$, we have $\rho$ related to $u$ and $r$ in the form of
$\rho\sim u^2/Gr^2$. In reference to equations (\ref{appen_x}) and
(\ref{appen_u}), we then have the mass density $\rho
=v^2(x)/(Gt^2x^2)$. Therefore, we put the factor $1/x^2$ into the
scale-free part and multiply a constant factor $1/(4\pi)$ for
notational convenience and arrive at
\begin{equation}
\rho=D(t)\alpha(x)=\alpha(x)/(4\pi Gt^2)\ .\label{appen_rho}
\end{equation}
There implies a relationship $\alpha(x)\propto v^2(x)/x^2$.

From the scaling relationship $M\sim\rho r^3$ and the two
self-similar expressions (\ref{appen_x}) and (\ref{appen_rho}), we
have the enclosed mass $M$ in the form of $M=k^{3/2}t^{3n-2}
x^3\alpha(x)/4\pi G$. We then merge the factor $x^3$ into the
scale-free part and multiply a numerical factor $4\pi/(3n-2)$ for
notational convenience. It follows that the enclosed mass is in
the form of
\begin{equation}
M=B(t)m(x)=k^{3/2}t^{3n-2}m(x)/[(3n-2)G]\ .\label{appen_m}
\end{equation}
It is also apparent that the scaling $m(x)\sim x^3\alpha(x)$
holds.

The thermal gas pressure $p$ has the scaling of $p\sim c^2
\rho\sim kt^{2n-4}/(4\pi G)$, where the form of the polytropic
sound speed $c\propto k^{1/2}t^{n-1}$ and equation
(\ref{appen_rho}) are used. For the scale-free part, we refer to
the equation of state (\ref{Equ:state}) and obtain the relation
$p=K(t)\rho^{\gamma}\sim\alpha^{\gamma}(x)$. Therefore, the
thermal gas pressure takes the following form
\begin{equation}
p=E(t)\beta(x)=kt^{2n-4}\alpha^{\gamma}(x)/(4\pi G)\ .
\label{appen_m}
\end{equation}
It immediately follows that $\beta(x)\sim \alpha^{\gamma}(x)$.

\section[]{Derivation of Self-Similar Ordinary Differential Equations}

Here we provide the derivation of the two nonlinear coupled ODEs
(\ref{Equ:function1}) and (\ref{Equ:function2}) using the
self-similar transformation (\ref{equ:varu}).

By the chain rule, the partial differentials of a general function
${\cal F}(r,t)=F(t)f(x)$ are:
\begin{eqnarray}
\frac{\partial{\cal F}(r,t)}{\partial r}
=F(t)\frac{df(x)}{dx}\frac{\partial x}{\partial r}
=\frac{F(t)}{k^{1/2}t^n}\frac{df(x)}{dx}\ ,\label{appen_partial1}
\end{eqnarray}
\begin{eqnarray}
\frac{\partial{\cal F}(r,t)}{\partial t}
=\frac{dF(t)}{dt}f(x)+F(t)\frac{df(x)}{dx}
\frac{\partial x}{\partial t} \qquad\nonumber\\
\quad =\frac{dF(t)}{dt}f(x)-\frac{nr F(t)}
{k^{1/2}t^{n+1}}\frac{df(x)}{dx}\ . \label{appen_partial2}
\end{eqnarray}
Referring to these partial differentials and the self-similar
transformation (\ref{equ:varu}), equations (\ref{Equ:mass1}),
(\ref{Equ:force}) and (\ref{Equ:mass2}) can be readily reduced to
\begin{equation}
(v-nx)\frac{d\alpha}{dx}+\alpha\frac{dv}{dx} +\frac{2\alpha
v}{x}-2\alpha=0\ ,\label{appen_b1}
\end{equation}
\begin{equation}
(n-1)v+(v-nx)\frac{dv}{dx}+\gamma\alpha^{\gamma-2}
\frac{d\alpha}{dx}+\frac{m}{(3n-2)x^2}=0\ , \label{appen_b2}
\end{equation}
\begin{equation}
(3n-2)m=(nx-v)\frac{dm}{dx}\ , \quad
\frac{dm}{dx}=(3n-2)x^2\alpha\ , \label{appen_b3}
\end{equation}
respectively. From equation (\ref{appen_b3}), it follows that
\begin{equation}
m(x)=\alpha x^2(nx-v)\ ,\label{appen_mres}
\end{equation}
leading to a necessary physical requirement of $nx-v>0$ for a
positive enclosed mass; this is referred to in Section 2 of the
main text.

Finally, from equations (\ref{appen_b1}), (\ref{appen_b2}) and
(\ref{appen_mres}), we derive the two coupled nonlinear ODEs in
terms of $v(x)$ and $\alpha(x)$
\begin{eqnarray}
{\big[(v-nx)^2-\gamma\alpha^{\gamma-1}\big](dv/dx)}
={(v-nx)^2}\alpha/{(3n-2)}\quad\nonumber\\
+2(v-x)\gamma\alpha^{\gamma-1}/x-(v-nx)(n-1)v\
\qquad\label{Equ:appen_function1}
\end{eqnarray}
\vskip -0.1cm \noindent and
\begin{eqnarray}
\big[(v-nx)^2-\gamma\alpha^{\gamma-1}\big] \big[d\alpha/(\alpha
dx)\big]=
(n-1)v \qquad\qquad\nonumber\\
-(v-nx)\alpha /(3n-2)-2(v-x)(v-nx)/x\ .
\qquad\label{Equ:appen_function2}
\end{eqnarray}

\section[]{Derivation of Shock Conditions
in the Self-Similar Framework}

We summarize here the derivation of shock conditions in the
self-similar framework, namely, equations (\ref{shock1}) and
(\ref{shock2}) in the main text.

In the reference framework of the shock front, the shock jump
conditions for the mass density and flow velocity are shown in
equation (\ref{shock_v}), where subscripts 1 and 2 denote the
pre-shock (upstream) and post-shock (downstream) sides,
respectively. Note that flow velocity in the reference framework
of the shock front has the form of $\bar{u}_1=u_1-u_s$ and
$\bar{u}_2=u_2-u_s$, where $u_s$ is the radial travel speed of the
shock front. The shock front can be identified with $x_1$ in the
pre-shock (upstream) side. Thus referring to equation
(\ref{equ:varx}), the travel speed of a shock front is
\begin{equation}
u_s=dr/dt=d(x_1k_1^{1/2}t^n)/dt=nk_1^{1/2}t^{n-1}x_1\ .
\label{shockvelocity1}
\end{equation}
This travel speed of the shock front can be also derived from the
post-shock (downstream) side, namely
\begin{equation}
u_s=dr/dt=d(x_2k_2^{1/2}t^n)/dt=nk_2^{1/2}t^{n-1}x_2\ .
\label{shockvelocity2}
\end{equation}
The shock front travel speed derived from the upstream and
downstream sides should be the same. Thus, from expressions
(\ref{shockvelocity1}) and (\ref{shockvelocity2}), we derive the
discontinuity in constant $k$, namely
\begin{equation}
(k_2/k_1)^{1/2}=x_1/x_2 \label{shock_k}\ .
\end{equation}
This discontinuity in $k$ originates physically from the
discontinuity in temperature or entropy across the shock front, as
discussed in Section 2.3 of the main text. We also list flow
velocities in the shock reference framework here
\begin{equation}
\bar{u}_1=u_1-u_s=k_1^{1/2}t^{n-1}(v_1-nx_1)\
\label{equ:velocity1}
\end{equation}
and
\begin{equation}
\bar{u}_2=u_2-u_s=k_2^{1/2}t^{n-1}(v_2-nx_2)\
.\label{equ:velocity2}
\end{equation}

Referring to the sound speed definition (\ref{equ:soundspeed}),
the pre-shock (upstream) sound speed is
$c_1=\gamma^{1/2}k_1^{1/2}t^{n-1}\alpha_1^{(\gamma-1)/2}$. Then
from equations (\ref{equ:mach}) and (\ref{equ:velocity1}), the
pre-shock (upstream) Mach number is
\begin{equation}
\bar{\cal
M}_1=\frac{\bar{u}_1}{c_1}=\frac{k_1^{1/2}t^{n-1}(v_1-nx_1)}
{k_1^{1/2}t^{n-1}\gamma^{1/2}\alpha_1^{(\gamma-1)/2}}
=\frac{v_1-nx_1}{\gamma^{1/2}\alpha_1^{(\gamma-1)/2}}\ ,
\label{equ:mach1}
\end{equation}
where $\bar{\cal M}$ is the Mach number in the reference framework
of the shock front. With this derived upstream Mach number, we
obtain from equations (\ref{equ:varu}) and (\ref{shock_v}) the
density discontinuity in the self-similar framework as
\begin{equation}
\alpha_2=\frac{\alpha_1(\gamma+1)(v_1-nx_1)^2}
{(\gamma-1)(v_1-nx_1)^2+2\gamma\alpha_1^{\gamma-1}}\ .
\label{equ:shock1C}
\end{equation}
Finally from equations (\ref{shock_v}),
(\ref{shock_k})$-$(\ref{equ:mach1}), we derive the radial flow
velocity discontinuity in the self-similar framework as
\begin{equation}
v_2=nx_2+\frac{x_2}{x_1(\gamma+1)}\bigg[(\gamma-1)
(v_1-nx_1)+\frac{2\gamma\alpha_1^{\gamma-1}}{v_1-nx_1}\bigg]\ .
\label{equ:shock2C}
\end{equation}

\end{document}